\pdfoutput=1

\documentclass[12pt,a4paper]{article}
  
\usepackage{ifthen} 
\newboolean{pdflatex}
\setboolean{pdflatex}{true} 

\newboolean{articletitles}
\setboolean{articletitles}{true} 

\newboolean{uprightparticles}
\setboolean{uprightparticles}{false} 

\newboolean{inbibliography}
\setboolean{inbibliography}{true} 
 
 \usepackage{overpic}

\newcommand{\fnot}{\ensuremath{f_0(980)}}
\newcommand{\kstar}{\ensuremath{K^{*}(892)^0}}

\newcommand{\rhor}{\ensuremath{\rho(770)^0}}
\newcommand{\pipi}{\ensuremath{\pi^+\pi^-}}
\newcommand{\jpsipipi}{\ensuremath{\jpsi\,\pipi}}
\newcommand{\mumu}{\ensuremath{\mu^+\mu^-}}
\newcommand{\mumupipi}{\ensuremath{\pipi\mumu}}
\newcommand{\pipimumu}{\mumupipi}
\newcommand{\Kpi}{\ensuremath{K^+\pi^-}}
\newcommand{\br}{\ensuremath{\mathcal{B}}}
\newcommand{\Rf}{\ensuremath{\mathcal{R}_{s}}}
\newcommand{\Rr}{\ensuremath{\mathcal{R}_{d}}}
\newcommand{\Bcand}{\ensuremath{B^0_{(s)}}}

\newcommand{\Bdpipimumu}{\ensuremath{{\decay{\Bd}{ \mumupipi }}}}
\newcommand{\Bspipimumu}{\ensuremath{{ \decay{\Bs }{ \mumupipi }}}}
\newcommand{\BdJpsipipi}{\ensuremath{{\decay{\Bd }{  \jpsipipi }}}}
\newcommand{\BsJpsipipi}{\ensuremath{{ \decay{\Bs }{ \jpsipipi }}}}
\newcommand{\Bdrhomumu}{\ensuremath{{ \decay{\Bd }{ \rhor \mumu }}}}
\newcommand{\Bsfnotmumu}{\ensuremath{{ \decay{\Bs }{ \fnot \mumu }}}}

\newcommand{\Bpipimumu}{\ensuremath{{\decay{B_{(s)}^0 }{  \mumupipi }}}}
\newcommand{\BJpsipipi}{\ensuremath{{\decay{B_{(s)}^0 }{  \jpsipipi }}}}
\newcommand{\BdJpsiKst}{\ensuremath{{\decay{\Bd }{ \jpsi \kstar }}}}
\newcommand{\BdKstmumu}{\ensuremath{{\decay{\Bd }{ \kstar \mumu }}}}
\newcommand{\BcJpsipipipi}{\ensuremath{{\decay{\Bc}{\jpsi \pipi\pi^+}}}}

\def\signBs{\ensuremath{7.2\sigma}}
\def\signBd{\ensuremath{4.8\sigma}}
\def\RfFinal{\ensuremath{  (1.67\pm 0.29\,\stat \pm 0.13\,\syst)\times 10^{-3}}}
\def\RrFinal{\ensuremath{ (0.41 \pm 0.10\,\stat \pm 0.03\,\syst)\times 10^{-3}}}

\def\BrBsFinal{\ensuremath{(8.6\pm 1.5\,\stat \pm 0.7\,\syst\pm 0.7\,({\rm norm}))\times 10^{-8}}}
\def\BrBdFinal{\ensuremath{(2.11\pm 0.51\,\stat \pm 0.15\,\syst\pm 0.16\,({\rm norm}) )\times 10^{-8}}}

\def\BrBsTrue{\ensuremath{(8.3\pm 1.7)\times 10^{-8}}}
\def\BrBdTrue{\ensuremath{(1.98\pm 0.53)\times 10^{-8}}}
 
\def\numBs{\ensuremath{55\pm 10\,\stat \pm 5\,\syst}}
\def\numBd{\ensuremath{40\pm 10\,\stat \pm 3\,\syst}}

\usepackage{microtype}
\usepackage{lineno}  
\usepackage{xspace} 
\usepackage{caption} 

\usepackage{graphicx}  
\usepackage{color}
\usepackage{colortbl}
\graphicspath{{./}} 

\usepackage{amsmath} 
\usepackage{amssymb}
\usepackage{amsfonts}
\usepackage{upgreek} 

\newcommand*\patchAmsMathEnvironmentForLineno[1]{%
\expandafter\let\csname old#1\expandafter\endcsname\csname #1\endcsname
\expandafter\let\csname oldend#1\expandafter\endcsname\csname
end#1\endcsname
 \renewenvironment{#1}%
   {\linenomath\csname old#1\endcsname}%
   {\csname oldend#1\endcsname\endlinenomath}%
}
\newcommand*\patchBothAmsMathEnvironmentsForLineno[1]{%
  \patchAmsMathEnvironmentForLineno{#1}%
  \patchAmsMathEnvironmentForLineno{#1*}%
}
\AtBeginDocument{%
\patchBothAmsMathEnvironmentsForLineno{equation}%
\patchBothAmsMathEnvironmentsForLineno{align}%
\patchBothAmsMathEnvironmentsForLineno{flalign}%
\patchBothAmsMathEnvironmentsForLineno{alignat}%
\patchBothAmsMathEnvironmentsForLineno{gather}%
\patchBothAmsMathEnvironmentsForLineno{multline}%
\patchBothAmsMathEnvironmentsForLineno{eqnarray}%
}

\usepackage{hyperref}    
\usepackage[all]{hypcap} 

\def\lhcb {\mbox{LHCb}\xspace}
\def\ux85 {\mbox{UX85}\xspace}

\ifthenelse{\boolean{uprightparticles}}%
{

 \def\Pmu         {\ensuremath{\upmu}\xspace}

 \def\Ppi         {\ensuremath{\uppi}\xspace}

 \def\Ppsi        {\ensuremath{\uppsi}\xspace}

 \def\PDelta      {\ensuremath{\Delta}\xspace}                 
 \def\PXi      {\ensuremath{\Xi}\xspace}                 
 \def\PLambda      {\ensuremath{\Lambda}\xspace}                 
 \def\PSigma      {\ensuremath{\Sigma}\xspace}                 
 \def\POmega      {\ensuremath{\Omega}\xspace}                 
 \def\PUpsilon      {\ensuremath{\Upsilon}\xspace}                 
 
 \def\PB      {\ensuremath{\mathrm{B}}\xspace}                 
                  
 \def\PD      {\ensuremath{\mathrm{D}}\xspace}

 \def\PJ      {\ensuremath{\mathrm{J}}\xspace}                 
 \def\PK      {\ensuremath{\mathrm{K}}\xspace}

 \def\Pb      {\ensuremath{\mathrm{b}}\xspace}                 
 \def\Pc      {\ensuremath{\mathrm{c}}\xspace}

 \def\Pi      {\ensuremath{\mathrm{i}}\xspace}

 \def\Ps      {\ensuremath{\mathrm{s}}\xspace}

}
{

 \def\Pmu         {\ensuremath{\mu}\xspace}

 \def\Ppi         {\ensuremath{\pi}\xspace}

 \def\Ppsi        {\ensuremath{\psi}\xspace}                 
                  
 \mathchardef\PDelta="7101
 \mathchardef\PXi="7104
 \mathchardef\PLambda="7103
 \mathchardef\PSigma="7106
 \mathchardef\POmega="710A
 \mathchardef\PUpsilon="7107
                  
 \def\PB      {\ensuremath{B}\xspace}                 
                  
 \def\PD      {\ensuremath{D}\xspace}

 \def\PJ      {\ensuremath{J}\xspace}                 
 \def\PK      {\ensuremath{K}\xspace}

 \def\Pb      {\ensuremath{b}\xspace}                 
 \def\Pc      {\ensuremath{c}\xspace}

 \def\Pi      {\ensuremath{i}\xspace}

 \def\Ps      {\ensuremath{s}\xspace}

}

\def\mumu       {\ensuremath{\Pmu^+\Pmu^-}\xspace}

\def\squark    {\ensuremath{\Ps}\xspace}

\def\cquark    {\ensuremath{\Pc}\xspace}

\def\bquark    {\ensuremath{\Pb}\xspace}

\def\pion  {\ensuremath{\Ppi}\xspace}

\def\pipi  {\ensuremath{\pion^+\pion^-}\xspace}

\def\kaon  {\ensuremath{\PK}\xspace}
  \def\Kbar  {\kern 0.2em\overline{\kern -0.2em \PK}{}\xspace}

\def\Kz    {\ensuremath{\kaon^0}\xspace}
\def\Kzb   {\ensuremath{\Kbar^0}\xspace}
\def\KzKzb {\ensuremath{\Kz \kern -0.16em \Kzb}\xspace}
\def\Kp    {\ensuremath{\kaon^+}\xspace}
\def\Km    {\ensuremath{\kaon^-}\xspace}

\def\KpKm  {\ensuremath{\Kp \kern -0.16em \Km}\xspace}

  \def\Dbar    {\kern 0.2em\overline{\kern -0.2em \PD}{}\xspace}
\def\D       {\ensuremath{\PD}\xspace}

\def\Dz      {\ensuremath{\D^0}\xspace}
\def\Dzb     {\ensuremath{\Dbar^0}\xspace}
\def\DzDzb   {\ensuremath{\Dz {\kern -0.16em \Dzb}}\xspace}
\def\Dp      {\ensuremath{\D^+}\xspace}
\def\Dm      {\ensuremath{\D^-}\xspace}

\def\DpDm    {\ensuremath{\Dp {\kern -0.16em \Dm}}\xspace}

\def\B       {\ensuremath{\PB}\xspace}
  \def\Bbar    {\kern 0.18em\overline{\kern -0.18em \PB}{}\xspace}

\def\Bz      {\ensuremath{\B^0}\xspace}

\def\Bu      {\ensuremath{\B^+}\xspace}

\def\Bd      {\ensuremath{\B^0}\xspace}
\def\Bs      {\ensuremath{\B^0_\squark}\xspace}

\def\Bc      {\ensuremath{\B_\cquark^+}\xspace}

\def\jpsi     {\ensuremath{{\PJ\mskip -3mu/\mskip -2mu\Ppsi\mskip 2mu}}\xspace}

  \def\Y#1S{\ensuremath{\PUpsilon{(#1S)}}\xspace}

\def\Lbar {\ensuremath{\kern 0.1em\overline{\kern -0.1em\PLambda}}\xspace}

\newcommand{\decay}[2]{\ensuremath{#1\!\to #2}\xspace}         

\def\to                 {\ensuremath{\rightarrow}\xspace}

\def\AT#1     {\ensuremath{A_{\mathrm{T}}^{#1}}\xspace}           

\def\C#1      {\ensuremath{\mathcal{C}_{#1}}\xspace}                       
\def\Cp#1     {\ensuremath{\mathcal{C}_{#1}^{'}}\xspace}                    
\def\Ceff#1   {\ensuremath{\mathcal{C}_{#1}^{\mathrm{(eff)}}}\xspace}        
\def\Cpeff#1  {\ensuremath{\mathcal{C}_{#1}^{'\mathrm{(eff)}}}\xspace}       
\def\Ope#1    {\ensuremath{\mathcal{O}_{#1}}\xspace}                       
\def\Opep#1   {\ensuremath{\mathcal{O}_{#1}^{'}}\xspace}                    

\newcommand{\tev}{\ensuremath{\mathrm{\,Te\kern -0.1em V}}\xspace}
\newcommand{\gev}{\ensuremath{\mathrm{\,Ge\kern -0.1em V}}\xspace}
\newcommand{\mev}{\ensuremath{\mathrm{\,Me\kern -0.1em V}}\xspace}
\newcommand{\kev}{\ensuremath{\mathrm{\,ke\kern -0.1em V}}\xspace}
\newcommand{\ev}{\ensuremath{\mathrm{\,e\kern -0.1em V}}\xspace}
\newcommand{\gevc}{\ensuremath{{\mathrm{\,Ge\kern -0.1em V\!/}c}}\xspace}
\newcommand{\mevc}{\ensuremath{{\mathrm{\,Me\kern -0.1em V\!/}c}}\xspace}
\newcommand{\gevcc}{\ensuremath{{\mathrm{\,Ge\kern -0.1em V\!/}c^2}}\xspace}
\newcommand{\gevgevcccc}{\ensuremath{{\mathrm{\,Ge\kern -0.1em V^2\!/}c^4}}\xspace}
\newcommand{\mevcc}{\ensuremath{{\mathrm{\,Me\kern -0.1em V\!/}c^2}}\xspace}

\def\mum  {\ensuremath{\,\upmu\rm m}\xspace}

\def\invfb   {\ensuremath{\mbox{\,fb}^{-1}}\xspace}

\newcommand{\stat}{\ensuremath{\mathrm{(stat)}}\xspace}
\newcommand{\syst}{\ensuremath{\mathrm{(syst)}}\xspace}

\newcommand{\chisq}{\ensuremath{\chi^2}\xspace}

\def\gsim{{~\raise.15em\hbox{$>$}\kern-.85em
          \lower.35em\hbox{$\sim$}~}\xspace}
\def\lsim{{~\raise.15em\hbox{$<$}\kern-.85em
          \lower.35em\hbox{$\sim$}~}\xspace}

\def\sPlot{\mbox{\em sPlot}}

\def\pt         {\mbox{$p_{\rm T}$}\xspace}

\def\evtgen     {\mbox{\textsc{EvtGen}}\xspace}
\def\pythia     {\mbox{\textsc{Pythia}}\xspace}

\def\geant      {\mbox{\textsc{Geant4}}\xspace}

\def\photos     {\mbox{\textsc{Photos}}\xspace}

\def\tell1  {TELL1\xspace}
\def\ukl1   {UKL1\xspace}

\newcommand{\ie}{\mbox{\itshape i.e.}}

\usepackage{cite} 
\usepackage{mciteplus}

\usepackage{booktabs} 

\begin{document}

\renewcommand{\thefootnote}{\fnsymbol{footnote}}
\setcounter{footnote}{1}


\begin{titlepage}
\pagenumbering{roman}

\vspace*{-1.5cm}
\centerline{\large EUROPEAN ORGANIZATION FOR NUCLEAR RESEARCH (CERN)}
\vspace*{1.5cm}
\hspace*{-0.5cm}
\begin{tabular*}{\linewidth}{lc@{\extracolsep{\fill}}r}
\ifthenelse{\boolean{pdflatex}}
{\vspace*{-2.7cm}\mbox{\!\!\!\includegraphics[width=.14\textwidth]{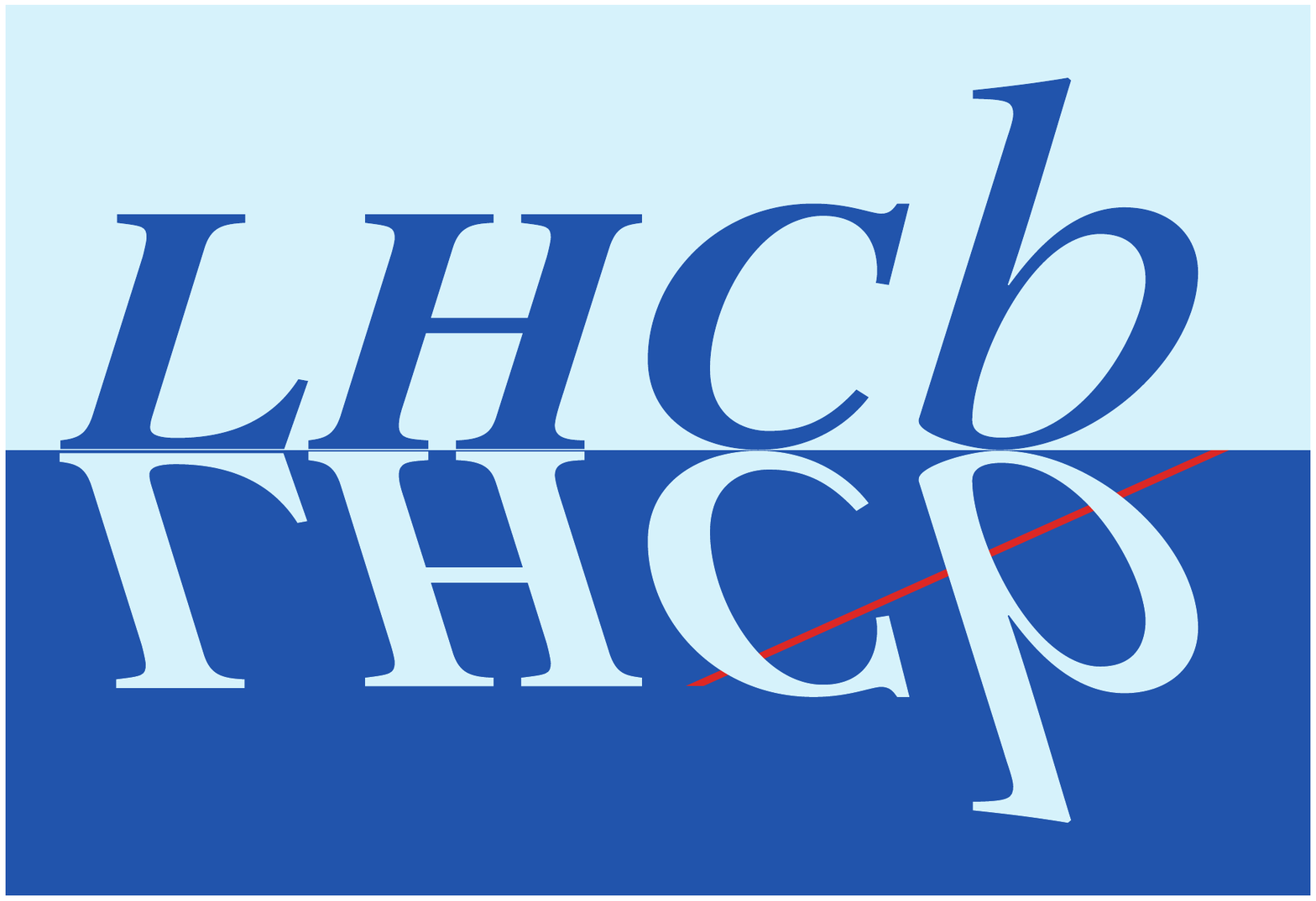}} & &}%
{\vspace*{-1.2cm}\mbox{\!\!\!\includegraphics[width=.12\textwidth]{lhcb-logo.eps}} & &}%
\\
 & & CERN-PH-EP-2014-296 \\  
 & & LHCb-PAPER-2014-063 \\  
 & & December 19, 2014 \\ 
 & & \\
\end{tabular*}

\vspace*{4.0cm}

{\bf\boldmath\huge
\begin{center}
  Study of the rare \Bs\ and \Bd\ decays into the \pipimumu\ final state 
\end{center}
}

\vspace*{1.0cm}

\begin{center}
The LHCb collaboration\footnote{Authors are listed at the end of this letter.}
\end{center}

\vspace{\fill}

\begin{abstract}
  \noindent
  A search for the rare decays  \Bspipimumu\ and \Bdpipimumu\  is performed in a data set corresponding 
  to an integrated luminosity of 3.0\invfb\ collected by the LHCb detector in proton-proton 
  collisions at centre-of-mass energies of 7 and 8\tev.  
   Decay candidates with pion pairs that have invariant mass in the range 0.5--1.3\gevcc 
  and with muon pairs that do not originate from a resonance are considered. 
   The first observation of the decay \Bspipimumu\ and the first evidence of the decay \Bdpipimumu\
    are obtained and the branching fractions, restricted to the dipion-mass range considered, 
    are measured to be
   $\br(\Bspipimumu)=\BrBsFinal$ and 
 $\br(\Bdpipimumu)=\BrBdFinal$, where the third uncertainty is due to 
  the branching fraction of the decay ${\decay{\Bd}{\decay{\jpsi(}{\mumu)}\decay{\kstar(}{\Kpi)}}}$, used as a normalisation. 
\end{abstract}

\vspace*{1.0cm}

\begin{center}
  Published as Phys. Lett. B743 (2015) 46 
\end{center}

\vspace{\fill}

{\footnotesize 
\centerline{\copyright~CERN on behalf of the \lhcb collaboration, licence \href{http://creativecommons.org/licenses/by/4.0/}{CC-BY-4.0}.}}
\vspace*{2mm}

\end{titlepage}


\newpage
\setcounter{page}{2}
\mbox{~}

\cleardoublepage


\renewcommand{\thefootnote}{\arabic{footnote}}
\setcounter{footnote}{0}



\pagestyle{plain} 
\setcounter{page}{1}
\pagenumbering{arabic}


\section{Introduction}
\label{sec:Introduction}
Decays of the \Bs\ and \Bd\ mesons into a 
\pipimumu\ final state  with the muons not originating from a resonance 
are  flavour-changing neutral-current transitions,\footnote{The inclusion of charge-conjugate processes is implied throughout.}    
which are expected to proceed mainly from the ${\decay{\Bs}{\decay{\fnot(}{\pipi)}\mumu}}$ and 
${\decay{\Bd}{\decay{\rhor(}{\pipi)}\mumu}}$ decays, 
in analogy to what is observed in ${\decay{\Bcand}{\jpsi\pipi}}$ 
decays~\cite{LHCb-PAPER-2014-012,LHCb-PAPER-2013-069}.  
In the standard model (SM) these decays are governed by the ${\decay{b}{s}}$ 
and ${\decay{b}{d}}$ weak transitions and are described by
 loop diagrams. They are suppressed due to the 
Glashow-Iliopoulos-Maiani mechanism~\cite{PhysRevD.2.1285} and  
the small values of the Cabibbo-Kobayashi-Maskawa  
matrix elements involved~\cite{PhysRevLett.10.531,Kobayashi01021973}. 
This feature makes the \Bsfnotmumu\ and \Bdrhomumu\ decays 
sensitive probes of several SM extensions, since potential non-SM amplitudes may dominate over the SM 
contribution~\cite{4q,extradim,Iltan-1999,Aliev-1999,b2rhomumu-2008}. 
 Current SM predictions of the \Bsfnotmumu\ 
branching fraction vary from $10^{-7}$ to $10^{-9}$~\cite{pQCD,Colangelo2010,Ghahramany:2009zz};
similar values are expected for the \Bdrhomumu\ branching 
fraction~\cite{KrugerSehgal1997,mel,Beneke:2004dp}. 
The predictions suffer from uncertainties in the calculation of the hadronic matrix elements  
 associated with the transitions. 
 For the \Bsfnotmumu\ decay, the limited knowledge of the quark content of the \fnot\  meson   
results in additional uncertainties.  
No experimental information exists on these decays to date. 

In this Letter, a search for the \Bpipimumu\ decays is reported.  
The analysis is restricted to events with muons that do not originate from 
$\phi$, \jpsi, and $\psi(2S)$ resonances,
and with pion pairs with invariant mass in the range 0.5--1.3\gevcc. 
This mass range is set to include both \fnot\ and \rhor\ resonances, 
which overlap because of their large widths~\cite{PDG2014}.  
Other resonances, as well as non-resonant pions, might contribute~\cite{LHCb-PAPER-2014-012,LHCb-PAPER-2013-069}. 
However, due to the limited size of the data sample, an amplitude analysis of the \pipi\ mass spectrum is not attempted.
The analysis is performed in a data set corresponding to 
an integrated luminosity of 3.0\invfb, collected by the LHCb detector
in proton-proton ($pp$) collisions. The first 1.0\invfb of data 
was collected in 2011 with collisions at the centre-of-mass energy of 7\tev; 
the remaining 2.0\invfb in 2012 at 8\tev. 
The signal yields are obtained from a 
fit to the unbinned \mumupipi\ mass distribution of the decay candidates. 
The fit modelling and the methods for the background estimation are validated on data, 
by fitting the \mumupipi\ mass distribution of  \BJpsipipi\ decays, while  
the branching fractions of \Bpipimumu\ decays are normalised using 
\BdJpsiKst\ decays reconstructed in the same data set.

\section{Detector and simulation}
\label{sec:detector}
The \lhcb detector~\cite{Alves:2008zz} is a single-arm forward
spectrometer covering the \mbox{pseudorapidity} range $2<\eta <5$,
designed for the study of particles containing \bquark or \cquark
quarks. The detector includes a high-precision tracking system
consisting of a silicon-strip vertex detector surrounding the $pp$
interaction region~\cite{LHCb-DP-2014-001}, a large-area silicon-strip detector located
upstream of a dipole magnet with a bending power of about
$4{\rm\,Tm}$, and three stations of silicon-strip detectors and straw
drift tubes~\cite{LHCb-DP-2013-003} placed downstream of the magnet.
The tracking system provides a measurement of momentum  with
a relative uncertainty that varies from 0.4\% at low momentum to 0.6\% at 100\gevc.
The minimum distance of a track to a primary vertex~(PV), the impact parameter~(IP), 
is measured with a resolution of 20\mum for charged particles with high transverse momentum (\pt).
Different types of charged hadrons are distinguished using information
from two ring-imaging Cherenkov detectors (RICH)~\cite{LHCb-DP-2012-003}. 
Photon, electron and hadron candidates are identified by a calorimeter system consisting of
scintillating-pad and preshower detectors, an electromagnetic
calorimeter and a hadronic calorimeter. Muons are identified 
by a system composed of 
 alternating layers of iron and multiwire
proportional chambers~\cite{LHCb-DP-2012-002}.

Samples of simulated events are used to determine the efficiency of selecting
\Bpipimumu\ and \BdJpsiKst\ decays, and  to study backgrounds. 
In the simulation, $pp$ collisions are generated using
\pythia~\cite{Sjostrand:2006za,*Sjostrand:2007gs} 
 with a specific \lhcb configuration~\cite{LHCb-PROC-2010-056}.  Decays of hadronic particles
are described by \evtgen~\cite{Lange:2001uf}, in which final-state
radiation is generated using \photos~\cite{Golonka:2005pn}. 
The model of Refs.~\cite{Balakireva:2009kn,Popov:2010zz,Colangelo2010} 
is used to describe \Bpipimumu\ decays. 
The interaction of the generated particles with the detector and its
response are implemented using the \geant
toolkit~\cite{Allison:2006ve, *Agostinelli:2002hh} as described in
Ref.~\cite{LHCb-PROC-2011-006}.

\section{Event selection}
\label{sec:selection}
The online event-selection (trigger) consists of a hardware stage,
 based on information from the calorimeter and muon
systems, followed by a software stage, which applies a full event
reconstruction~\cite{LHCb-DP-2012-004}. 
For this analysis, the hardware trigger requires  
at least one muon with ${\pt>1.48\,(1.76)\gevc}$, or 
two muons with ${\sqrt{\pt(\mu_1)\,\pt(\mu_2)}>1.3\,(1.6)\gevc}$, in the 2011 (2012) data sample. In
the software trigger, at least
one of the final-state particles is required to have 
$\pt>1\gevc$ and $\rm{IP}>100\mum$ with respect to all
 the primary $pp$ interaction vertices in the
event. Finally, the tracks of two or more  final-state
particles are required to form a vertex that is significantly
displaced from the PVs. A multivariate algorithm 
is used to identify  
secondary vertices consistent with the decay 
of a $b$ hadron~\cite{2013JInst8P2013G}. 

In the offline selection,  
all charged particles are required to have $\pt>0.25$\gevc 
and trajectories not consistent with originating from the PVs. 
Two oppositely charged muon candidates compatible with originating from the same displaced 
vertex are considered. To reject ${\decay{\phi}{ \mumu}}$, ${\decay{ \jpsi}{\mumu}}$, 
and ${\decay{ \psi(2S)}{\mumu}}$ decays, candidates having invariant mass in the ranges  
1.010--1.030, 2.796--3.216, or 3.436--3.806\gevcc are removed;  
 contributions from other resonances 
in the \mumu\ mass spectrum such as $\rhor$, $\omega(782)$, and $\psi(4160)$~\cite{LHCb-PAPER-2013-039} are negligible. 
The muon candidates are combined with a pair of oppositely charged pions   
with invariant mass in the range 0.5--1.3\gevcc to form \Bpipimumu\ candidates.  
For the ${\decay{\Bd}{\decay{\jpsi(}{\mumu)} \decay{\kstar(}{\Kpi)}}}$ candidates, 
the dimuon invariant mass is required to be in the range 2.796--3.216\gevcc, and the
invariant mass of the pion and kaon system in the range 0.826--0.966\gevcc.   
The four tracks are required to originate from the same \Bcand\ decay vertex. 
The \Bcand\ momentum vector is required to be within 14\,mrad of the vector
that joins the PV with the \Bcand\ decay vertex (flight distance vector).

The information from the RICH, the calorimeters, 
and the muon systems is used for particle identification (PID), \ie, 
to define a likelihood for each track to be  
associated with a certain particle hypothesis. 
Requirements on the muon-identification likelihood 
 are applied to reduce to $\mathcal{O}(10^{-2})$ 
the rate of misidentified muon candidates, mainly pions, 
whilst preserving 95\% signal efficiency.  
In the case of \BdJpsiKst\ decays, PID requirements on kaon candidates are applied  
to suppress any contributions from \BJpsipipi\ decays with pions misidentified
 as kaons. In the case of \Bpipimumu decays, a requirement 
 on the PID of pion candidates is applied to reduce the 
contamination from $\decay{\Bd}{\decay{\kstar(}{\Kpi)}\mumu}$ decays with kaons misidentified 
as pions; this background peaks around 5.25\gevcc
in the \mumupipi\ mass spectrum. A large data set of \BdJpsipipi\ decays is used 
to optimise the PID requirement of pion candidates, 
assuming that the proportion between misidentified \BdJpsiKst\ 
 and \BdJpsipipi\  decays is similar to the proportion between misidentified \BdKstmumu\  
 and \Bdpipimumu\ decays.   
The requirement retains about 55\% of the signal candidates. 
Simulations show that additional contributions from ${\decay{\Bs}{\decay{\phi(}{ K^+ K^-)}\mumu}}$ decays 
with double kaon-pion misidentification are negligible. 
A requirement on the proton-identification likelihood of pion candidates  
suppresses the contamination from decays with protons misidentified as pions, 
 with  a $95\%$ signal efficiency.   
After this selection, simulations show that contributions 
from ${\decay{\Lambda_b^0}{\decay{\Lambda(}{ p\pi^-)} \mumu}}$ and 
${\decay{\Lambda_b^0}{ p\pi^- \mumu}}$ decays are negligible, as are    
contributions from ${\decay{\Lambda_b^0}{\decay{ \Lambda(1520)(}{ pK^-)} \mumu}}$ and 
${\decay{\Lambda_b^0}{ pK^- \mumu}}$ decays, where both the  proton and 
 the kaon are misidentified as pions.

In addition to the above requirements, a multivariate selection based on 
a boosted decision tree~(BDT)~\cite{Breiman,AdaBoost} is used to 
suppress the large background from random combinations of tracks 
(combinatorial background) present in the \pipimumu\ sample. 
The BDT is trained using  simulated \Bspipimumu\ 
events to model the signal, and data candidates with \mumupipi\  
mass in the range 5.5--5.8\gevcc for the background. 
The training is performed separately for 
the 2011 and 2012 data, and using simulations that reproduce the specific  
operational conditions of each year. 
The variables used in the BDT are the significance of the displacement from the PV 
of pion and muon tracks, the fit \chisq\ of the \Bcand\ decay vertex,  
the angle between the \Bcand\ momentum vector and the 
flight distance vector, the \pt of the \Bcand\ candidate, 
 the sum and the difference of the transverse momenta of pions, the difference of 
 the transverse momenta of muons, 
the \Bcand\ decay time, and
the minimum \pt of the pions. The resulting BDT output is 
 independent of the \mumupipi\  mass
and PID variables. 
A requirement on the BDT output value is chosen to maximise 
the figure of merit $\varepsilon/(\alpha/2 + \sqrt{N_b})$ \cite{Punzi:2003bu}, 
 where $\varepsilon$ is the signal efficiency; 
 $N_b$ is the number of background events that pass the selection 
 and have a mass within 30\mevcc of the known value of the \Bs mass~\cite{PDG2014}; 
 $\alpha$ represents  the desired significance 
 of the signal, expressed in terms of number of 
 standard deviations. The value of 
 $\alpha$ is set to 3 (5) for the 2011 (2012) data set. 
 The resulting selection has around 85\% efficiency to select signal candidates. 
 The same BDT is used to select \BdJpsiKst\ candidates. 
 The selected samples consist of 364 \Bpipimumu\ candidates and 52\,960 \BdJpsiKst\ candidates. 
    
The efficiencies of all selection requirements are estimated with simulations, 
except for the efficiency of the PID selection for hadrons. The latter  
is determined in data using large and low-background samples of 
$D^{*+}\to D^0(\to K^-\pi^+)\pi^+$ decays;  
the efficiencies are evaluated after reweighting the calibration samples 
to match simultaneously the momentum and pseudorapidity distributions 
of the final-state particles of \Bpipimumu\ (\BdJpsiKst) candidates, 
and the distribution of the track multiplicity of the events.
The final selection efficiencies for 2011 and 2012 data 
 are reported in Table~\ref{tab:efficiencies}. 
The statistical uncertainties are due to the size of the calibration and simulation samples; 
systematic uncertainties are described in what follows.  
The total efficiency varies by approximately 15\%  in the \pipi\ mass 
range considered and it is parametrised with a second-order polynomial. 
The signal candidates are weighted in order to have a constant efficiency as a function of the \pipi\ mass spectrum. 
\begin{table}
\caption{\label{tab:efficiencies} Selection efficiencies of the 2011 and 2012 data sets;
$\varepsilon_s$ for the \Bspipimumu\ decay, 
$\varepsilon_d$ for the \Bdpipimumu\ decay, and
$\varepsilon_n$ for the \BdJpsiKst\ decay.}
\centering
\begin{tabular}{lcc}
\toprule
 & 2011 & 2012 \\
\midrule
$\varepsilon_s$ [\%]& $36.1 \pm  0.3\,\stat \pm 2.4\,\syst$ & $36.9  \pm  0.3\,\stat\pm 2.3\,\syst$ \\
$\varepsilon_d$ [\%]& $29.8  \pm  0.2\,\stat \pm 2.0\,\syst$ & $27.5  \pm  0.2\,\stat \pm 1.7\,\syst$ \\
$\varepsilon_n$ [\%]& $9.33 \pm  0.05 \,\stat \pm 0.35\,\syst$ & $9.74  \pm  0.08 \,\stat \pm 0.27\,\syst$\\
\bottomrule
\end{tabular}
\end{table}

Systematic uncertainties of the efficiencies  are dominated by the 
limited information about the signal decay-models; the main
 contribution comes from the unknown angular distributions of 
 \Bpipimumu\ decay products.  
To estimate this uncertainty, the difference in efficiencies 
between decays generated according to a phase-space model 
and to the model of Refs.~\cite{Balakireva:2009kn,Popov:2010zz,Colangelo2010} is considered. 
The resulting relative uncertainty is $5.4\%$. 
A relative uncertainty of $3.7\%$ ($2.8\%$) for 2011 (2012) data 
is estimated by considering the difference of the efficiencies evaluated 
in the simulation and in data for \BdJpsiKst\ decays. 
The same  relative uncertainty is assigned to the efficiency associated with \Bpipimumu\ decays,  
as the cancellation of this uncertainty in the ratio of 
the efficiencies of signal and normalisation
decays may not be exact. This is due to the fact that the \pt\ distributions of the 
final-state particles are different between the decay modes.
An additional $1.6\%$ relative uncertainty is assigned to $\varepsilon_s$, 
due to the unknown mixture of \Bs\ mass 
eigenstates in \Bspipimumu\ decays, which results in a \Bs\ effective lifetime 
that could differ from the value used in the simulations~\cite{DeBruyn:2012wj}.

\section{Determination of the signal yields}
\label{sec:fit}
The ratio of the branching fractions   
\begin{equation}
\mathcal{R}_q \equiv \frac{ \br(\Bpipimumu) }{ \br(\decay{\Bd}{\decay{\jpsi(}{\mumu)}\decay{\kstar(}{\Kpi)}}) }, \nonumber
\end{equation}
with $q=s$ ($d$) for \Bspipimumu\ (\Bdpipimumu) decays, is the quantity being measured; 
it is used to express the observed yields of \Bpipimumu\ decays as follows:
\begin{equation}
N_{B_q} =  \frac{f_q}{f_d} \frac{ \varepsilon_q}{ \varepsilon_n}  N_n \mathcal{R}_q,  \label{eq:signal_yields1}
\end{equation} 
where $N_n$ is the \BdJpsiKst\ yield,  $f_s/f_d$ is the ratio of the fragmentation 
probabilities for \Bs\ and \Bd\ mesons~\cite{fsfd},  
$\varepsilon_q$ is the selection efficiency of \Bspipimumu\ (\Bdpipimumu) decays, and
$\varepsilon_n$ the one of \BdJpsiKst\ decays.   

\begin{figure}[tb]
\center
\includegraphics[width=0.60\textwidth]{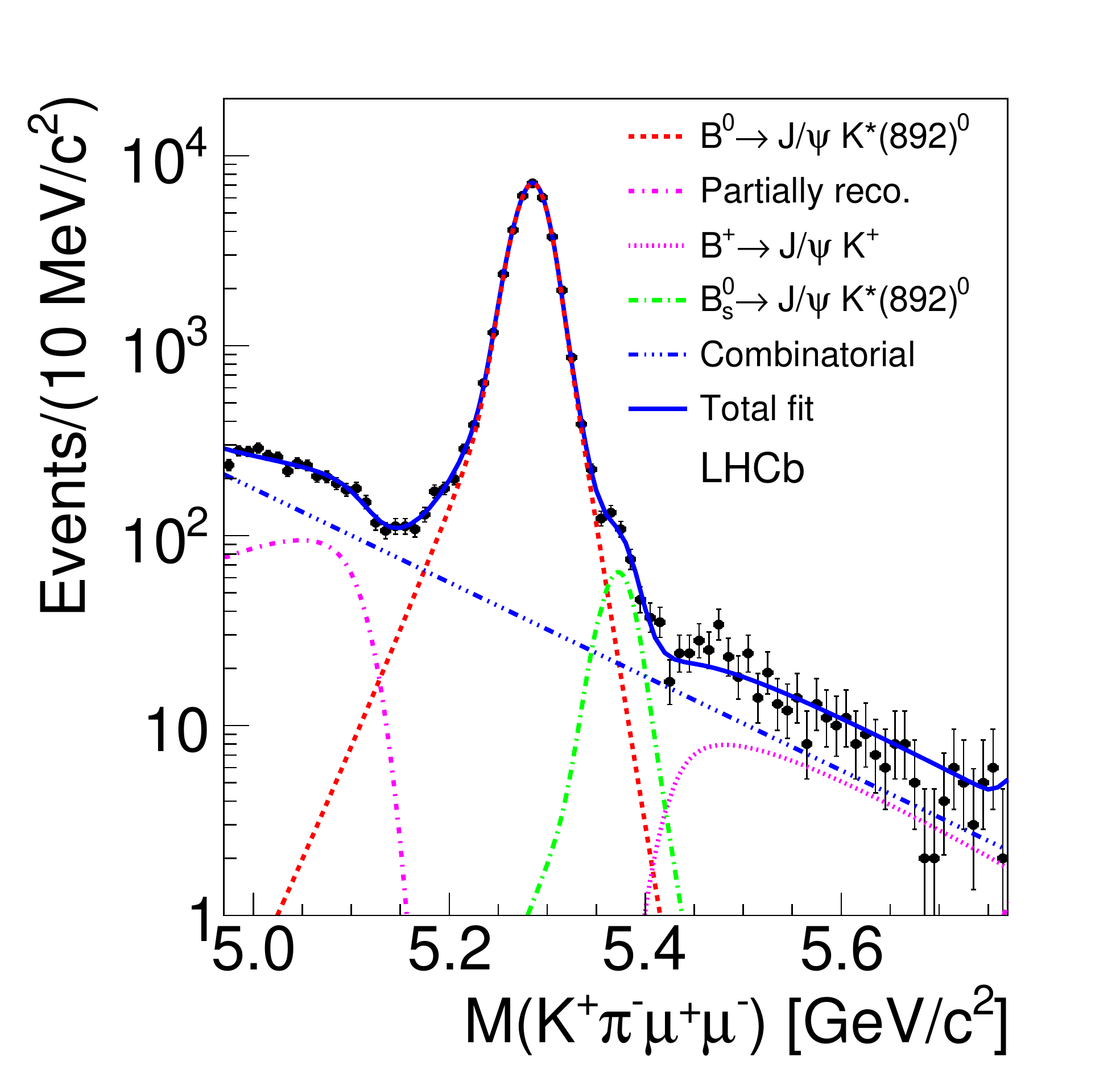} 
\caption{Mass distribution of \BdJpsiKst\ candidates with fit 
projections overlaid. The 2011 and 2012 data sets are combined.}
\label{fig:fit_BdJpsiKst}
\end{figure}
The number of events $N_n$ in Eq.~(\ref{eq:signal_yields1}) is obtained from an extended maximum likelihood 
fit to the unbinned $\mumu\Kpi$ mass distribution of the \BdJpsiKst\ candidates 
in the range 4.97--5.77\gevcc. 
The $\mumu\Kpi$ mass distribution is shown in Fig.~\ref{fig:fit_BdJpsiKst} with fit projections overlaid. 
A sum of two Gaussian functions, with a power-law tail on either side 
 derived from simulations, is used to describe the
dominant \BdJpsiKst\ peak and the small $\Bs\to\jpsi\kstar$ contribution.  
All function parameters are in common between the 
 \Bz\ and \Bs\ signal functions, except for the mass; 
 the mass difference  between \Bs\ and \Bz\ mesons is fixed to the known value~\cite{PDG2014}.
An exponential function is used to model the combinatorial background. 
A small contamination of $\Bu\to\jpsi K^+$ decays combined with an additional charged pion is 
modelled with an ARGUS function~\cite{Albrecht1989304}. Partially reconstructed \Bz\ 
decays at masses lower than the \Bz\ signal  
are described with another ARGUS function.  
The fitted yields of \BdJpsiKst\ decays are corrected by subtracting a $(6.4\pm1.0)\%$ contribution of 
${\decay{\Bz}{\jpsi\Kpi}}$ decays~\cite{LHCb-PAPER-2013-023}, where the $\Kpi$ pair is in a $S$-wave state  
and does not originate from the decay of a $\kstar$ resonance.  
The numbers of \BdJpsiKst\ decays are  $9821\pm110\,\stat\pm134\,\syst \pm97\,(S\rm{\,wave})$  
and $23521\pm175\,\stat\pm172\,\syst\pm243\,(S\rm{\,wave})$ in the 2011 and 2012 data sets,
 respectively,  where the 
 third uncertainty is due to the $S$-wave subtraction. 
The systematic uncertainty accounts for the uncertainties in the parameters 
fixed in the fit to the values determined in simulations, and 
  are calculated with the method described at the end of this section. 

\begin{figure}[tb]
\center
\begin{overpic}[width=0.49\textwidth]{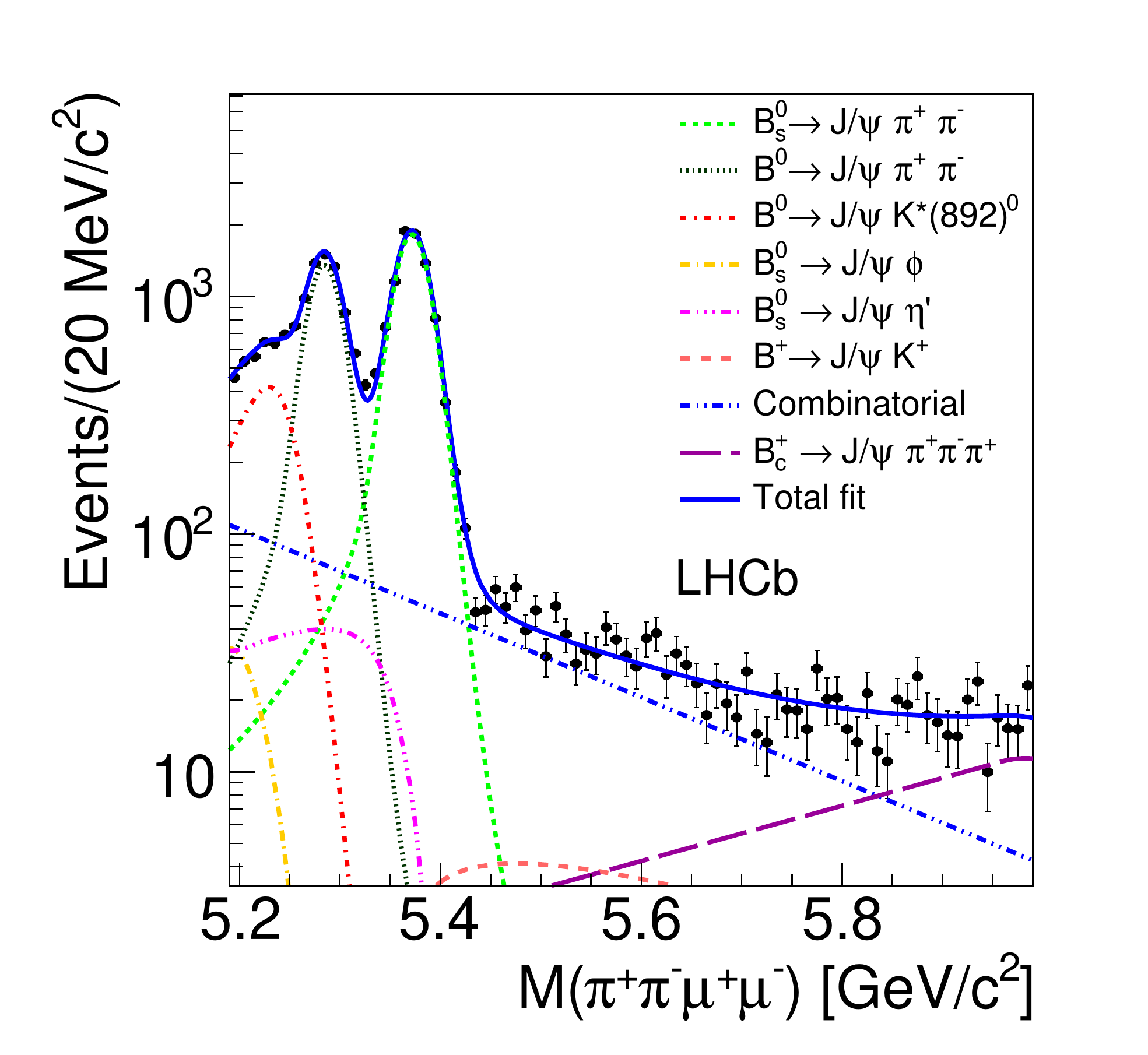} 
\put(24,78){(a)}
\end{overpic}
\begin{overpic}[width=0.49\textwidth]{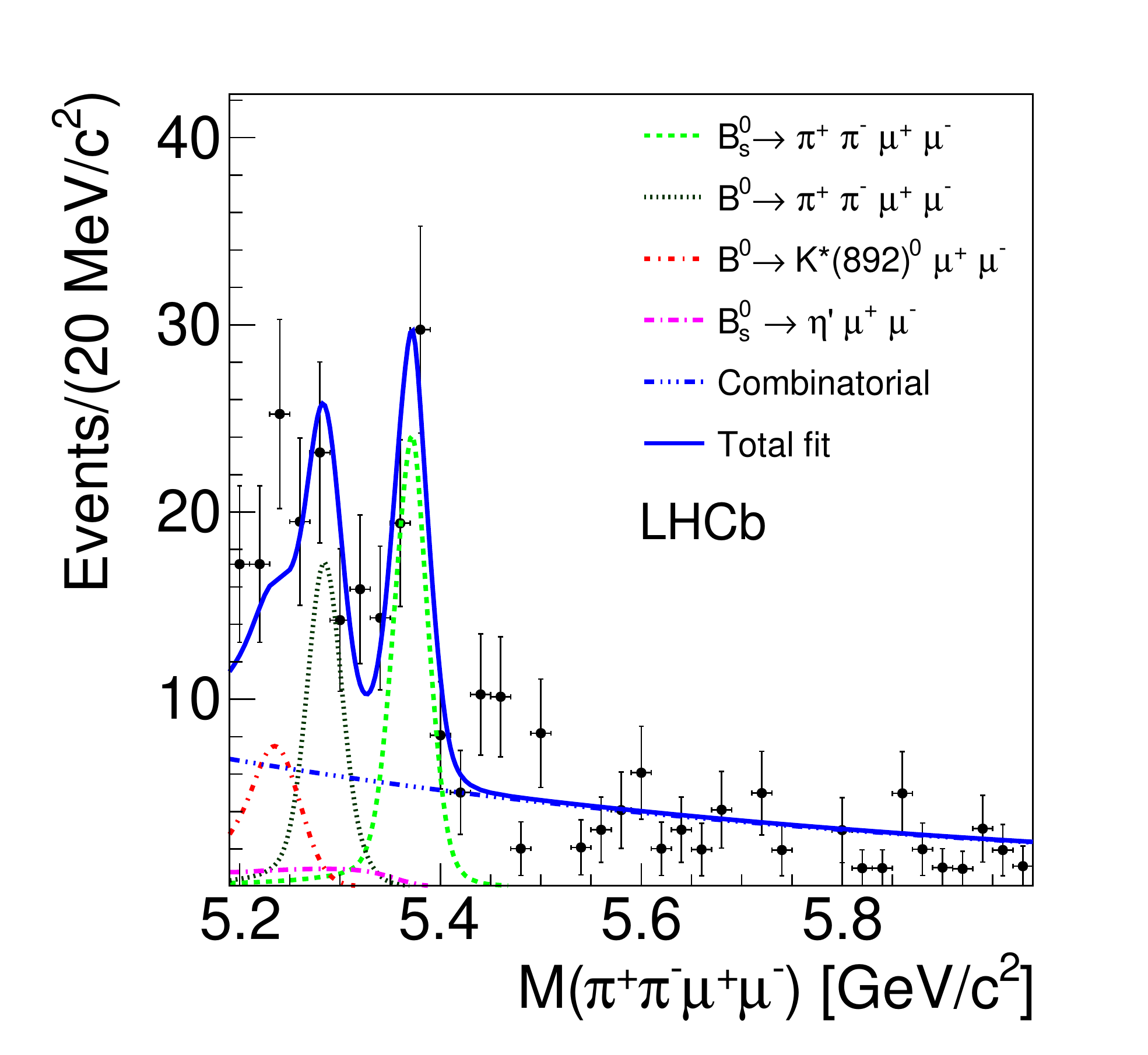}
\put(24,78){(b)}
\end{overpic}
\caption{Mass distributions of (a) the \BJpsipipi\ and  (b) the \Bpipimumu\
decay candidates in the range 5.19--5.99\gevcc\ with fit projections overlaid. 
The 2011 and 2012 data sets are combined. In (b), the contribution
from ${\decay{\Bs}{\phi \mumu}}$ and ${\decay{\Bu}{K^+ \mumu}}$ decays are included in the fit,  
but they are not visible in the projection, because the corresponding yields are small.}
\label{fig:fit_mumupipi}
\end{figure}

The ratios \Rf\ and \Rr\ are measured from an  extended maximum likelihood  fit to the
unbinned \mumupipi\  mass distribution, where the signal yields are
parametrised using Eq.~(\ref{eq:signal_yields1}), 
and all other inputs are fixed. 
The different centre-of-mass energies   
result in different $b\bar{b}$ production cross sections and 
selection efficiencies in the 
2011 and 2012 data samples. Therefore, the two samples are fitted 
simultaneously with different likelihood functions, but with the parameters \Rf\ and \Rr\ in common.  
We also fit simultaneously the \Bpipimumu\ and \BJpsipipi\ samples. 
The latter are selected with the \Bpipimumu\ requirements, 
except for the dimuon  mass, which is restricted to the 2.796--3.216\gevcc\ range.
The \BJpsipipi\ fit  serves as a consistency check of the fit modelling, since the \Bpipimumu\ and 
 \BJpsipipi\ mass distributions are expected to be similar.  
In both samples, the fit range is 2\gevcc wide and starts from 5.19\gevcc. 
This limit is set to remove partially reconstructed decays of the \Bd\ mesons with an unreconstructed~$\pi^0$.  
The stability of the fit results is checked against the extension
of the fit range in the lower mass region of the \Bpipimumu\ and \BJpsipipi\
 mass distributions, where an additional component is needed
in the fit to describe the partially reconstructed \Bd\ decays below 5.19\gevcc.
 Figure~\ref{fig:fit_mumupipi} shows the \mumupipi\ 
mass distributions of the \BJpsipipi\ and \Bpipimumu\ decay 
candidates in the range 5.19--5.99\gevcc with fit projections overlaid, 
where the 2011 and 2012 data sets are combined. 
 
The \Bpipimumu\ and \BJpsipipi\ signals are described 
by a model similar to that used for the \BdJpsiKst\ signal in the fit of the \mumu\Kpi\ mass distribution.  
The \Bd\ peak position is a common parameter for the \Bpipimumu\ and \BJpsipipi\ fits, 
as well as the signal resolutions; the difference between the \Bd\ and the \Bs\ masses is fixed to the known value. 
The \Bpipimumu\ signal widths  
 are multiplied by scale factors, derived from simulations, 
which accounts for the different momentum spectra between non-resonant muons and muons from  
\jpsi\ meson decays.  In both fits, the combinatorial background is modelled with an exponential function.

Backgrounds from \BdKstmumu\ (\BdJpsiKst) decays,
where kaons are misidentified as pions, are estimated using 
control samples of these decays reconstructed in data. 
They are selected as \Bpipimumu\ (\BJpsipipi) candidates, except for different  
requirements on the PID variables of the kaon and pion candidates, 
as for the normalization decay mode. 
To obtain the yields and the shapes of the mass distribution of the misidentified decays,  
the kaon candidates are assigned the pion mass, and 
the resulting \mumupipi\ mass distribution is reweighted 
 to reproduce the PID selection of the \Bpipimumu\ sample. 
In the final fit, the  yields of the two backgrounds are constrained using Gaussian functions 
with means fixed to the values obtained
with this method, and widths that account for a 
relative uncertainty in the 2011 (2012) data sample of 
15\% (10\%) for \BdKstmumu\ decays, 
and of 2\% (1\%) for \BdJpsiKst\ decays.
The shape of the \BdKstmumu\ background is modelled
with a Gaussian function with a power-law tail on the low-mass side;
the shape of the \BdJpsiKst\ background  is modelled 
with a sum of two Gaussian functions with different means.  
All parameters of these functions are fixed from the values obtained 
in the fit to the control samples. 
The background from ${\decay{\Bs}{\jpsi\kstar}}$ 
decays is expected to be less than 0.5\%~\cite{PDG2014} of the
\BdJpsiKst\ yield and is neglected. Similarly, the background from ${\decay{\Bs}{\kstar\mumu}}$  decays 
is not considered. 
 
Backgrounds from decays ${\decay{\Bs}{ \decay{\phi (}{ \pi^+ \pi^- \pi^0)} \mumu}}$ 
with an unreconstructed~$\pi^0$, 
${\decay{\Bs}{ \decay{\eta' (}{  \pi^+ \pi^- \gamma)} \mumu}}$ 
with an unreconstructed~$\gamma$, and ${\decay{\Bu}{ K^+\mumu}}$ or ${\decay{\Bu}{\pi^+\mumu}}$  
 combined with an additional charged pion, are estimated from simulations. 
The mass distributions of these backgrounds  are modelled with ARGUS functions 
with parameters fixed from fits to simulated events.  
Backgrounds from similar decay modes, 
where the muons come from the \jpsi\ meson,
are described in the \BJpsipipi\ fit using the same methods. 
 An additional contribution is given by \BcJpsipipipi\ decays, 
where a pion is not reconstructed.  
 This background 
 is modelled with a sum of two Gaussian functions, one of which has a power-law tail on the low-mass side. 
Backgrounds from semileptonic   
${\decay{\Bd}{\decay{\D^-(}{\rho^0 \mu^- X)}\mu^+ X}}$ decays 
with ${\decay{\rho^0}{\pipi}}$, 
give a negligible contribution at \pipimumu\ mass greater than
 5.19\gevcc. 

\section{Results}
We measure $\Rf =  \RfFinal$ and 
$\Rr = \RrFinal$. Systematic uncertainties are discussed 
below. These values correspond to \numBs\ \Bspipimumu\ 
decays and \numBd\ \Bdpipimumu\ decays.  
The significances of the observed signals are calculated using Wilks' theorem~\cite{wilks1938}, and  
 are \signBs\ and \signBd\ for the \Bspipimumu\ and \Bdpipimumu\ decays, respectively. 
The \Bspipimumu\ (\Bdpipimumu) significance is obtained by considering the \Bdpipimumu\ (\Bspipimumu)
yield as a floating parameter in the fit. 
The systematic uncertainties are included by multiplying the significance by the factor
$1/\sqrt{1+(\sigma_{\syst}/\sigma_{\stat})^2}$, where  $\sigma_{\stat}$ 
is the statistical uncertainty, and $\sigma_{\syst}$ is the sum in quadrature 
of the contributions in Table~\ref{tab:systematics}, except for the uncertainty on $f_s/f_d$. 

\begin{figure}[t]
\center
\begin{overpic}[width=0.48\textwidth]{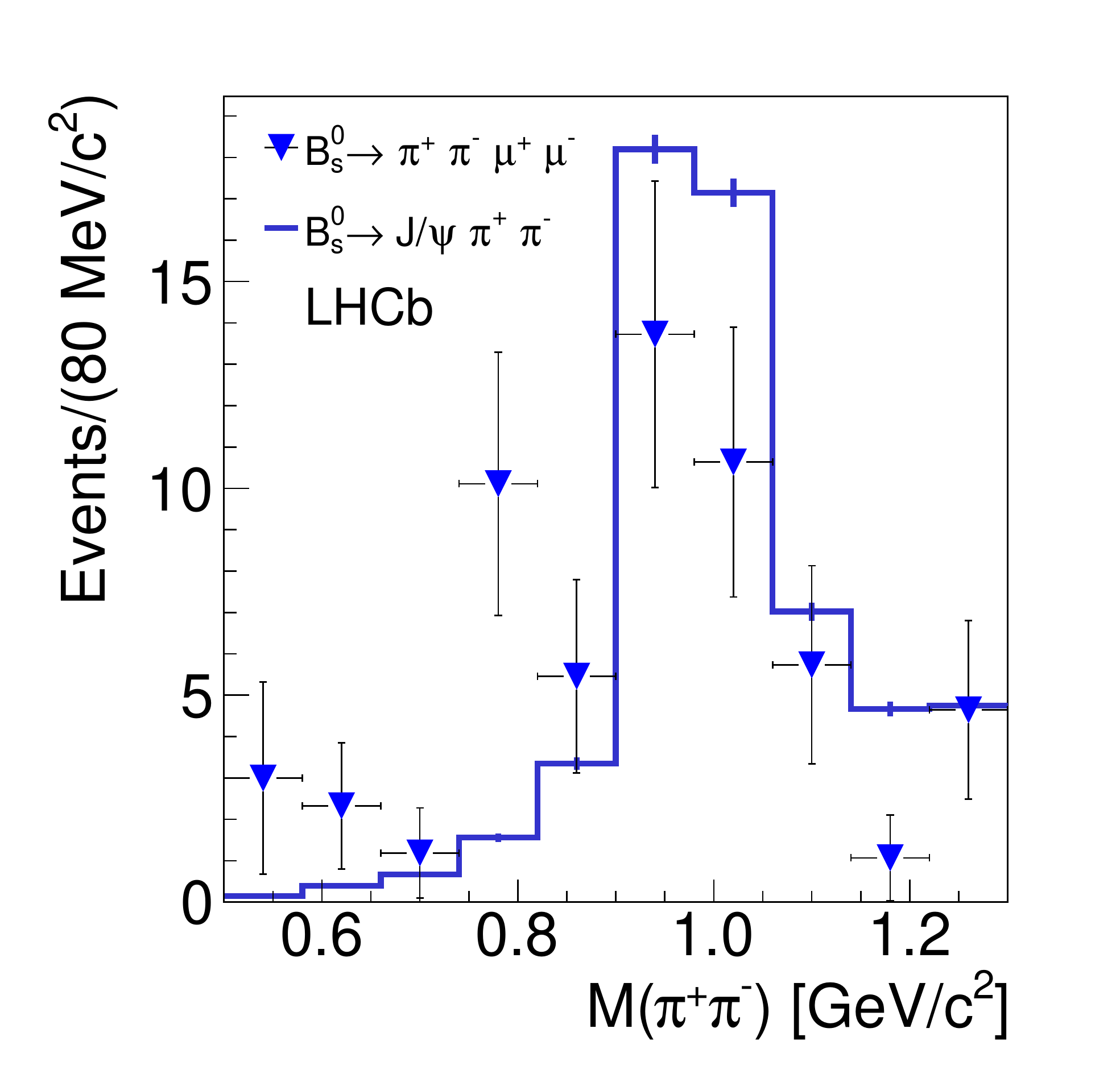} 
\put(78,80){(a)}
\end{overpic} \hfill
\begin{overpic}[width=0.48\textwidth]{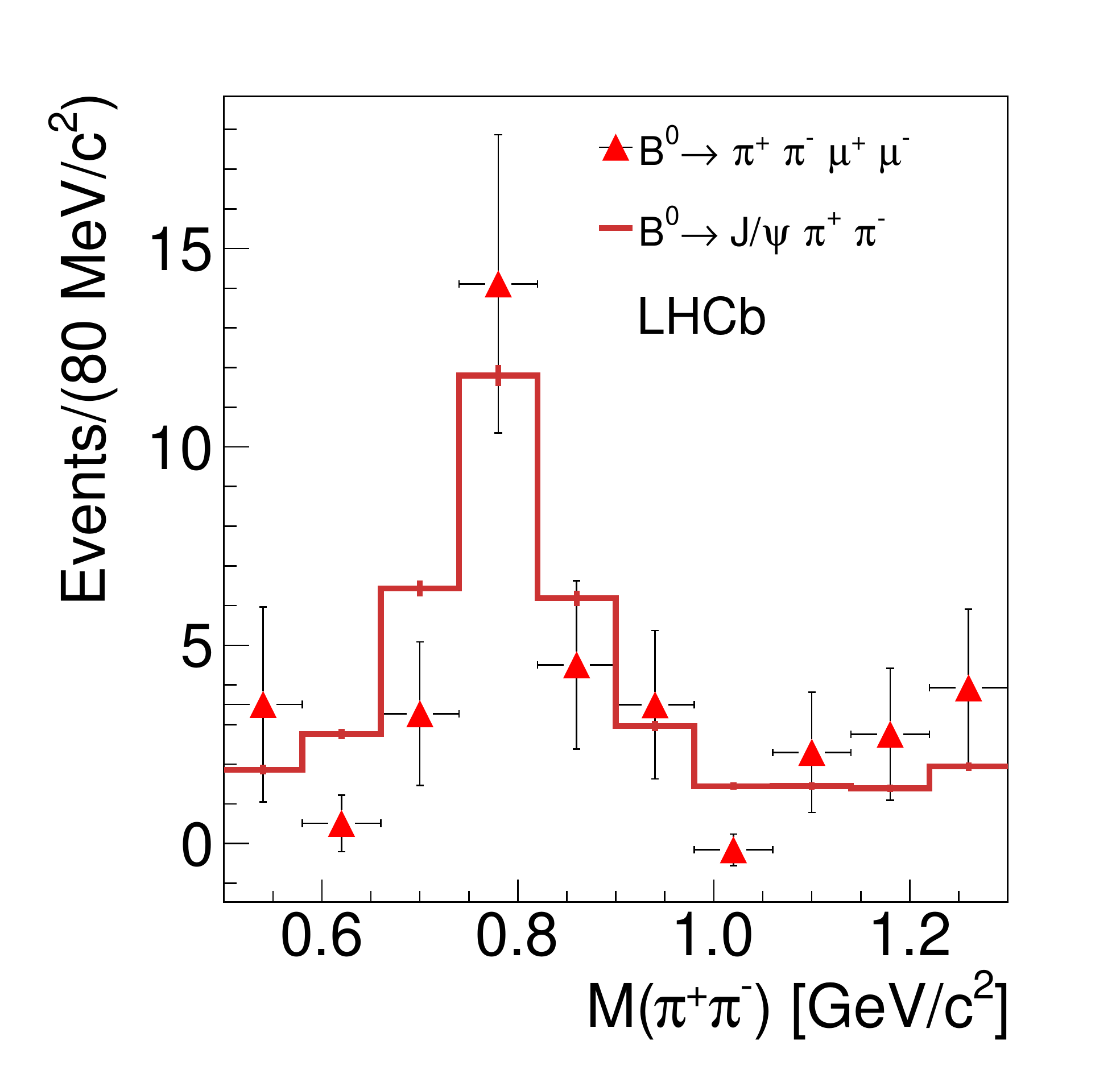} 
\put(25,80){(b)}
\end{overpic}
\caption{Background-subtracted distributions of the \pipi\ invariant mass for 
(a)~\Bspipimumu\ and (b)~\Bdpipimumu\ candidates (triangular markers). The uncertainties are
statistical only. 
The data are compared with the background-subtracted \pipi\ mass distributions of (a) \BsJpsipipi\ and (b) \BdJpsipipi\ 
candidates (histograms). }
\label{fig:pipi_mass}
\end{figure}

 Figure~\ref{fig:pipi_mass} compares the \pipi\ mass spectra of  
 \Bpipimumu\ and \BJpsipipi\ candidates, separately for the \Bs\ and the 
\Bz\ decays. The background is subtracted using the 
\sPlot\ technique~\cite{Pivk:2004ty} with 
 the \mumupipi\ mass  as the discriminating variable. 
 The data show the dominance of the \fnot\ resonance in the case of \BsJpsipipi\ decays, 
 and of the \rhor\ resonance in the case of \BdJpsipipi\ decays, as expected
 from previous LHCb analyses~\cite{LHCb-PAPER-2014-012,LHCb-PAPER-2013-069}. 
 The \Bpipimumu\ data show indications of a similar composition of the \pipi\ mass spectrum, 
 although the size of the sample is not sufficient to draw a definite conclusion. 
 
 Several systematic uncertainties on $\Rf$ and $\Rr$ are considered, as summarised in Table~\ref{tab:systematics}. 
The contribution due to the uncertainties on parameters that are fixed in the fit, and on the efficiencies and the yields of \BdJpsiKst\ decays that are fixed in Eq. (1), is obtained by repeating the fit, each time with the relevant parameters or inputs fixed to alternate values.
These are sampled from Gaussian distributions centred at the nominal value, and whose widths correspond to the uncertainties on the fixed parameters and inputs.  
Known correlations between fixed parameters are taken into account.  
 The r.m.s. spreads of the resulting \Rf\ and \Rr\ values are taken as the systematic uncertainties. 
The uncertainties associated with efficiencies are the sums in quadrature of their statistical and 
systematic uncertainties, reported in Table~\ref{tab:efficiencies}. 
The uncertainty on the \BdJpsiKst\ yield is the sum in quadrature of the statistical uncertainty, 
the systematic uncertainty, and the uncertainty due to the $S$-wave subtraction. 
A systematic uncertainty is assigned on the estimation of 
the combinatorial background with the following method; pseudo experiments 
are generate in an extended mass range from 4.97\gevcc, where an additional peaking
component is also added to simulate the partially reconstructed \Bd\ decays, and the pseudo data 
are fitted in the nominal range from 5.19\gevcc. The shifts between the average fitted values 
and the input values of \Rf\ and \Rr\ are taken as the systematic uncertainties.    
The contribution to the systematic uncertainty of \Rf\ due to the uncertainty on the values of 
$f_s/f_d$ is also included.  
The final systematic uncertainties are the sums in quadrature of all contributions and  
correspond to $45\%$ and $28\%$   
of the statistical uncertainties of \Rf\ and \Rr, respectively. 
\begin{table}[t]
\caption{Summary of systematic uncertainties on \Rf\ and \Rr.}
 \label{tab:systematics}
\begin{center}
\begin{tabular}{lcc}
\toprule  
      Source    & $\sigma(\Rf)$  $[10^{-3}]$       & $\sigma(\Rr)$   $[10^{-3}]$   \\ 
\midrule
Shape of misidentified decays &0.003& 0.004\\
Partially reconstructed decays  & 0.003& 0.004 \\
Combinatorial background & 0.029 & 0.014 \\
Signal shapes &0.020&0.014\\
Efficiencies &0.061&0.013\\
Normalisation decay yields & 0.055 & 0.014\\
$f_s/f_d$ &0.093& --\\
\midrule
Quadratic sum &0.130&0.028\\
\bottomrule
\end{tabular}
\end{center}
\end{table}

\section{Conclusions}
\label{sec:conclusions}
The first observation of the decay \Bspipimumu\ and the first evidence of the decay
  \Bdpipimumu\ are obtained in a data set corresponding to an 
   integrated luminosity of 3.0\invfb\ collected by the LHCb detector
  in $pp$  collisions at centre-of-mass energies of 7 and 8\tev. 
  The analysis is restricted to candidates with muon pairs  that do not originate from 
  	$\phi$, $\jpsi$, and $\psi(2S)$ resonances, while   
 the pion pairs are required to have invariant  mass in the range 0.5--1.3\gevcc. 
About 55 \Bspipimumu\ decays and 40 \Bdpipimumu\ decays are observed with 
significances of \signBs\ and  \signBd, respectively. 
Their branching fractions relative to the branching fraction of the 
${\decay{\Bd}{\decay{\jpsi(}{\mumu)}\decay{\kstar(}{\Kpi)}}}$ decay are
measured to be
\begin{align}
\frac{\br(\Bspipimumu)}{\br(\decay{\Bd}{\decay{\jpsi(}{\mumu)}\decay{\kstar(}{\Kpi)}})} &=\RfFinal, \nonumber \\
\frac{\br(\Bdpipimumu)}{\br(\decay{\Bd}{\decay{\jpsi(}{\mumu)}\decay{\kstar(}{\Kpi)}})} &=\RrFinal. \nonumber
\end{align}
From these ratios, the following branching fractions are obtained 
for the decays with the dipion-mass range considered: 
\begin{align}
\br(\Bspipimumu) &= \BrBsFinal \text{ and} \nonumber \\
\br(\Bdpipimumu) &= \BrBdFinal ,\nonumber 
\end{align}
where the third uncertainties are due to the uncertainties on 
the branching fraction of the normalization decay.   
We use $\br(\BdJpsiKst)=(1.30 \pm 0.10)\times 10^{-3}$, which is the weighted average of measurements 
where the \Kpi\ $S$-wave contribution is subtracted~\cite{Aubert:2007qea,Abe:2002haa,Jessop:1997jk},
 ${\br(\decay{\jpsi}{\mumu})}$ from Ref.~\cite{PDG2014}, and  ${\br(\decay{\kstar}{\Kpi})=2/3}$.
 
Assuming that the decays ${\fnot\to\pipi}$ and ${\rhor\to\pipi}$ are the dominant 
transitions in the \Bspipimumu\ and \Bdpipimumu\ decays, respectively, 
and neglecting other contributions, 
the \Bpipimumu\ branching fractions are corrected to account for the selection efficiencies of the \fnot\ and \rhor\ 
 resonances in the \pipi\ mass range considered. The following values 
 are obtained: $\br({\decay{\Bs}{\decay{\fnot(}{\pipi)}\mumu})=\BrBsTrue}$ and $\br(\Bdrhomumu)=\BrBdTrue$,
  where all uncertainties are summed in quadrature. 
These values favour SM expectations of Refs.~\cite{Colangelo2010,KrugerSehgal1997,mel} 
and disfavour the $\br(\Bsfnotmumu)$ SM expectation of Ref.~\cite{Ghahramany:2009zz}.

\section*{Acknowledgements} 
\noindent We express our gratitude to our colleagues in the CERN
accelerator departments for the excellent performance of the LHC. We
thank the technical and administrative staff at the LHCb
institutes. We acknowledge support from CERN and from the national
agencies: CAPES, CNPq, FAPERJ and FINEP (Brazil); NSFC (China);
CNRS/IN2P3 (France); BMBF, DFG, HGF and MPG (Germany); INFN (Italy); 
FOM and NWO (The Netherlands); MNiSW and NCN (Poland); MEN/IFA (Romania); 
MinES and FANO (Russia); MinECo (Spain); SNSF and SER (Switzerland); 
NASU (Ukraine); STFC (United Kingdom); NSF (USA).
The Tier1 computing centres are supported by IN2P3 (France), KIT and BMBF 
(Germany), INFN (Italy), NWO and SURF (The Netherlands), PIC (Spain), GridPP 
(United Kingdom).
We are indebted to the communities behind the multiple open 
source software packages on which we depend. We are also thankful for the 
computing resources and the access to software R\&D tools provided by Yandex LLC (Russia).
Individual groups or members have received support from 
EPLANET, Marie Sk\l{}odowska-Curie Actions and ERC (European Union), 
Conseil g\'{e}n\'{e}ral de Haute-Savoie, Labex ENIGMASS and OCEVU, 
R\'{e}gion Auvergne (France), RFBR (Russia), XuntaGal and GENCAT (Spain), Royal Society and Royal
Commission for the Exhibition of 1851 (United Kingdom).


\addcontentsline{toc}{section}{References}
\setboolean{inbibliography}{true}
\bibliographystyle{LHCb}
\bibliography{main,LHCb-PAPER,LHCb-CONF,LHCb-DP,LHCb-TDR}
   
 \pagebreak
 \newpage
 
\centerline{\large\bf LHCb collaboration}
\begin{flushleft}
\small
R.~Aaij$^{41}$, 
B.~Adeva$^{37}$, 
M.~Adinolfi$^{46}$, 
A.~Affolder$^{52}$, 
Z.~Ajaltouni$^{5}$, 
S.~Akar$^{6}$, 
J.~Albrecht$^{9}$, 
F.~Alessio$^{38}$, 
M.~Alexander$^{51}$, 
S.~Ali$^{41}$
% ...

{\footnotesize \it
% ...
$ ^{5}$Clermont Universit\'{e}, Universit\'{e} Blaise Pascal, CNRS/IN2P3, LPC, Clermont-Ferrand, France\\
$ ^{6}$CPPM, Aix-Marseille Universit\'{e}, CNRS/IN2P3, Marseille, France\\
% ...
$ ^{9}$Fakult\"{a}t Physik, Technische Universit\"{a}t Dortmund, Dortmund, Germany\\
% ...
$ ^{37}$Universidad de Santiago de Compostela, Santiago de Compostela, Spain\\
$ ^{38}$European Organization for Nuclear Research (CERN), Geneva, Switzerland\\
% ...
$ ^{41}$Nikhef National Institute for Subatomic Physics, Amsterdam, The Netherlands\\
% ...
$ ^{46}$H.H. Wills Physics Laboratory, University of Bristol, Bristol, United Kingdom\\
% ...
$ ^{51}$School of Physics and Astronomy, University of Glasgow, Glasgow, United Kingdom\\
$ ^{52}$Oliver Lodge Laboratory, University of Liverpool, Liverpool, United Kingdom\\
% ...
}
\end{flushleft}


\newpage



\end{document}